**Space-charge region recombination in monocrystalline silicon-based barrier structures with long lifetimes and its impact on key characteristics of high-efficiency solar cells**

A.V. Sachenko[1], V.P. Kostylyov[1], and M. Evstigneev[2]

[1] V. Lashkaryov Institute of Semiconductor Physics, NAS of Ukraine, 41 prospect Nauky, 03028 Kyiv, Ukraine

[2] Department of Physics and Physical Oceanography, Memorial University of Newfoundland, St. John's, NL, A1B 3X7 Canada

**Abstract.** The recombination rate in the space charge region (SCR) of a silicon-based barrier structure with long Shockley-Reed-Hall lifetime is calculated theoretically taking into account the concentration gradient of excess electron-hole pairs in the base region. The effects of the SCR lifetime and the applied voltage on the structure's ideality factor are analyzed. The ideality factor is significantly reduced by the concentration gradient of electron-hole pairs. This mechanism provides an increase of the effective lifetime compared to the case when it is insignificant, which is realized at sufficiently low pair concentrations. The theoretical results are shown to be in agreement with experimental data. A method of finding the experimental recombination rate in the SCR in highly efficient silicon solar cells (SCs) is proposed and implemented. It is shown that at high excess carrier concentration exceeding $10^{15}$ cm$^{-3}$ the contribution to the SCR recombination velocity from the initial region of the SCR that became neutral is significant. From a comparison of theory with experiment, the SCR life time and the ratio of the hole to the electron capture cross sections are determined for a number of silicon SCs. The effect of the SCR recombination on the key characteristics of high-efficiency silicon SCs, such as photoconversion efficiency and open-circuit voltage, is evaluated. It is shown to depend not only on the charge carrier lifetime in the SCR, but also on the ratio of hole to electron capture cross sections $\sigma_p/\sigma_n$. When $\sigma_p/\sigma_n < 1$, this effect is significantly strengthened, and in the opposite case $\sigma_p/\sigma_n > 1$ it is weakened. The effect described in the paper is also significant for silicon diodes with a thin base, p-i-n structures, and for silicon transistors with p-n junctions. In the appendix, the need to take into account the lifetime of non-radiative excitonic Auger recombination with the participation of deep impurities in silicon is analyzed in detail. It is shown, in particular, that its consideration makes it possible to reconcile the theoretical and experimental dependences for the effective life time in the silicon bulk.

# 1. Introduction

More than 60 years have passed since the publication of the classic study [1] of the recombination rate in the *p-n* junction space-charge region (SCR). After its publication, physicists actively published works in which various aspects of this problem were considered [2-7]. But even in the 21st century, this subject has not yet been fully exhausted (see, for example, Ref. [8]).

An approximate analysis of the generation-recombination processes in the depletion layer is given in the monograph [9]. It is based on the assumption that the quasi-Fermi levels of both charge carrier types are constant in the SCR. An examination of the expression for the Shockley-Reed-Hall (SRH) recombination rate shows that the maximum value of the recombination rate corresponds to the position where the intrinsic Fermi energy, $E_{Fi}$, is equally distant from the quasi-Fermi levels of electrons and holes. On both sides of the maximum, the recombination rate decreases exponentially with the characteristic length $kT/qE$, where $E$ is the electric field strength in the SCR, $q$ is the elementary charge, $k$ is Boltzmann's constant, $T$ is absolute temperature and $kT$ is the thermal energy. The effective thickness of the recombination region can be represented as $2kT/qE = kT\, w_d /q\, (V_d - V)$, where $w_d$ is the depletion region thickness, $V_d$ is the built-in voltage and $V$ is the applied voltage. From this analysis, the authors [9] conclude that for a spatially homogeneous distribution of recombination centers, the diode ideality factor $n$ is always less than two.

In the work [1] this issue was analyzed more thoroughly for a symmetrical p-n junction. It was established that the diode ideality factor is a function of the recombination center energy $E_t$, the temperature $T$, and the applied voltage $V$, and its maximum value is equal to 1.8 when $E_t \approx E_i$.

Choo [4] improved the theory [1] for the case of asymmetric junctions, in which the lifetimes of electrons $\tau_{n0}$ and holes $\tau_{p0}$ can vary in a wider range. As compared to the classical theory [1], the analysis performed in [4] gives smaller generation-recombination currents, which reach saturation at high forward bias. The values of the ideality factor can be both smaller and larger than 2.

In [10], a more complex and more accurate model was developed compared to the Sah-Noise-Shockley (SNS) model [1]. It differs from the SNS model in two respects. First, the potential $\psi(x)$ is found from the Poisson equation. Second, the depletion approximation is not used to determine the thickness $w_d$ of the SCR; rather, the integration limits were defined from the condition $d\psi(x)/dx = 0$.

In [11], an analytical method for calculating the recombination current in the depletion region of a forward-biased diode with a sharp transition is proposed. The method uses the Shockley–Read–Hall recombination approximation through one energy level, the concentration of which does not vary much depending on the position in the SCR. This model is systematically compared with previous models and with the results of numerical calculations using PC-1D software. It is shown that the proposed method [11], despite its simplicity, gives results closer to PC-1D numeric simulation than previous models. In addition, it is shown that the Nussbaum model [10] can be better agreed with the numerical results by reducing the integration limits by 0.3 $kT$.

The DESSIS semiconductor device simulation program was used in [12] to model the recombination current and determine the ideality factor. For this, a complete set of differential equations for semiconductors was solved without using the aforementioned approximations. Numerical modeling of DESSIS was used to determine the most accurate values of ideality factor and compare with them theoretical models, in particular [10,11]. At the same time, it was shown that the Nussbaum model [10] gives results sufficiently close to the results of numerical simulation.

It should be noted that the ideality factor of silicon diodes at low applied voltages is not always related to the SCR recombination. First, the initial section of the I-V cure is usually determined by the shunt resistance, so that the ideality coefficient in this part of the curve may significantly exceeds two in the range of voltages as broad as ca. 0.3 V. Secondly, in silicon diodes or pn-junction barrier structures with sufficiently long bulk lifetimes on the order of or greater than 1 ms, there is another physical mechanism that provides an ideality factor of 2. Namely, when the excess concentration $\Delta n$ in significantly exceeds the level of doping, i.e. $\Delta n \gg N_d$ (for definitness, we will consider an *n*-type semiconductor), the activation energy becomes comparable to half the bandgap, $E_a \approx E_g/2$, and the value of the ideality factor $n \approx 2$. This is exemplified by the heterostructure from [10], in which the ideality factor has the value of 1.8 at the applied voltage *V* ranging from 0.4 to 0.65 V.

In this work, the physical mechanism that ensures a reduction of the ideality factor compared to the value $n = 2$ in silicon *p-n*-junction structures with long bulk lifetimes of charge carriers is discussed. It is operative at sufficiently high recombination rates in the SCR and is related to the concentration gradient of excess electron-hole pairs in the bulk of the semiconductor. The condition for the emergence of this gradient is a large difference in the recombination velocities on the opposite different surfaces. Namely, on the surface with a *p-n* junction, the high net recombination rate is due to the SCR recombination, especially at low excitation levels. On the opposite surface which does not have the SCR, the total recombination rate is significantly lower. In its pure form, this mechanism manifests itself when the diffusion length significantly exceeds the thickness of the structure (more precisely, its base region) *d*. It is shown that this mechanism provides an increase in the effective lifetime of excess electron-hole pairs compared to the case when it is insignificant, which is realized at sufficiently low values of Δ*n*. A method is proposed to determine experimentally the recombination rate in the SCR in silicon $p^+$-$n$-$n^+$ structures. It is shown that at high excess carrier concentration exceeding $10^{15}$ cm$^{-3}$ the contribution to the SCR recombination velocity from the initial region of the SCR that became neutral is significant.

The classification of the effect of SCR recombination on the key characteristics of high-efficiency silicon SCs, depending on the minority charge carriers' lifetime $\tau_R$ in the SCR is performed. The results of works [13, 14] were used, in which it was established that the lifetime in the SCR is significantly shorter than the Shockley-Reed-Hall lifetime in the base region and can be of the order of or less than one microsecond.

A similar situation occurs during passivation of silicon with the SiN$_x$ layers (see, for example, [15]), when a significant positive static fixed charge gets built in into the dielectric. Then, the conductivity inversion occurs near the p-type silicon surface, and the SCR recombination becomes significant. In the work [15], lifetimes of the order of 1 microsecond were also observed in the SCR.

**2. Expressions for the SCR recombination velocity in the presence of excess carrier concentration gradient in the bulk semiconductor**

We will consider the case when the SCR recombination is determined by a single deep impurity level, whose filling is described by the Shockley-Reed-Hall statistics. Then, the value of the recombination velocity can be found by integrating the inverse recombination lifetime $\tau_R^{-1}(x)$ over the SCR thickness *w*:

$$S_{sc} = \int_0^w \frac{\tau_R^{-1}(x)(n_0 + \Delta n)dx}{(n_0 + \Delta n)e^{y(x)} + n_i(T)e^{E_t/kT} + b_r\left((p_0 + \Delta n)e^{-y(x)} + n_i(T)e^{-E_t/kT}\right)}. \tag{1}$$

Here $y(x)$ is the electric potential in the SCR divided by the thermal voltage $kT/q$, $n_0$ and $p_0$ are the equilibrium electron and hole concentrations, $p_0=n_i(T)^2/n_0$, $n_i(T)$ is the intrinsic concentration, $\Delta n$ is the excess carrier concentration and $b_r = C_p/C_n$ is the ratio of the hole and electron capture coefficients, which are expressible in terms of the respective thermal velocities, $V_{n,p}$, and capture cross-sections, $\sigma_{n,p}$, as $C_{n,p} = V_{n,p}\, \sigma_{n,p}$.

Almost all works devoted to the calculation of the recombination rate in the SCR analyze the case when the SCR recombination time $\tau_R(x) = $ const. Changing the integration variable from the coordinate $x$ to the dimensionless potential $y$, we obtain

$$S_{sc}(\Delta n) = \int_{y_w}^{y_0} \frac{\tau_R^{-1}(n_0 + \Delta n)\,dy}{(n_0 + \Delta n)e^y + n_i(T)\, e^{E_t/kT} + b_r\left((p_0 + \Delta n)e^{-y} + n_i(T)e^{-E_t/kT}\right)} F(y), \qquad (2)$$

where

$$F(y) = \frac{L_D}{\left(\dfrac{n_0 + \Delta n}{n_0}(e^y - 1) + y + \dfrac{p_0 + \Delta n}{n_0}(e^{-y} - 1)\right)^{1/2}}. \qquad (3)$$

Here $L_D = \left(\varepsilon_0 \varepsilon_{Si} kT / 2q^2 n_0\right)^{1/2}$ is Debye length, $q$ is the elementary charge, $y_0$ is the non-equilibrium non-dimensional band bending value on the surface of the weakly doped region, which depends on the injection level $\Delta n$ and is found from the integral neutrality condition, and $y_w$ is the non-equilibrium non-dimensional potential on the boundary between the SCR and the quasineutral region.

The dependence of the non-equilibrium dimensionless potential $y$ on the coordinate $x$ is found from the Poisson's equation:

$$x = \int_{y_0}^{y} \frac{L_D}{\left(\dfrac{n_0 + \Delta n}{n_0}(e^{y_1} - 1) - y_1 + \dfrac{\Delta n}{n_0}(e^{-y_1} - 1)\right)^{1/2}} dy_1. \qquad (4)$$

The value of the non-equilibrium dimensionless potential $y_0$ at $x = 0$ is found from the solution of the integral electric neutrality equation, which has the form

$$N = \pm \left(\frac{2kT\varepsilon_0 \varepsilon_{Si}}{q^2}\right)^{1/2} \left[(n_0 + \Delta n)(e^{y_0} - 1) - n_0 y_0 + \Delta n(e^{-y_0} - 1)\right], \qquad (5)$$

where $qN$ is the surface charge density of acceptors in the *p-n*-junction or in an anisotypic heterojunction.

The simplest expression for the recombination velocity in the SCR is obtained for the case of a small excitation level $\Delta n \ll n_0$, and the recombination level is located in the middle of the band gap. In this case, the integrand in (2) has a symmetric bell-shaped character and can be integrated over $y$ up to the maximum point $y_m = (1/2) \ln(n_0/(b_r(n_i(T)^2/n_0 + \Delta n))$. Then we get

$$S_{SC}^{an}(\Delta n) \approx \frac{kL_D}{\tau_R} \frac{\exp(y_m)}{\sqrt{y_m}}, \qquad (6)$$

where $k$ is a numerical coefficient of the order of 2.

Expression (6) can be used to calculate the value of $S_{SC}$ if $y_m > 2$. It should be noted that it does not apply near the point of maximum power collection of a silicon $p$-$n$ junction SC, because in this case the criterion $y_m > 2$ is not fulfilled. It breaks down especially early when $b_r > 1$. As for expression (2), it allows finding the recombination velocity in the SCR numerically at any ratio $\Delta n/n_0$.

This paper considers the case when the Shockley-Reed-Hall (SRH) lifetime in silicon is greater than 1 ms. In this case, the diffusion length significantly exceeds the thickness of the base region, i.e. the inequality $L \gg d$ holds, where $L = (D\tau_{eff}^v)^{1/2}$. Here $D$ is the diffusion coefficient of excess electron-hole pairs, and $\tau_{eff}^v$ is the effective lifetime of charge carriers in the base region. The excess concentration of charge carriers in the base is constant when the criterion $S_{sum} \ll D/d$ is fulfilled, where $S_{sum}$ is the total recombination rate in the SCR and on the front and rear surfaces of the base region of the semiconductor barrier structure of thickness $d$. If this criterion is not fulfilled due to the fact that the rate of recombination on one of the surfaces of the base is significantly greater than the rate of recombination on the second surface, then the $\Delta n$ value will depend on the $x$ coordinate. This case is realized at low levels of excitation precisely due to recombination in the SCR.

Let us first consider the situation when the SCR recombination occurs on the back surface at $x = d$. The generation-recombination balance equation in this case has the following form

$$J_{SC} = q\left[\int_0^d \frac{\Delta n(x)dx}{\tau_{eff}^b(x)} + S_0\Delta n(0) + S_{SC}\Delta n(d) + S_d\Delta n(d)\right], \qquad (7)$$

where $J_{SC}$ is the short-circuit current density, which is approximately equal to photocurrent $J_L$, $J_{SC} \cong J_L$, $\tau_{eff}^b$ is the effective bulk life time in the base, $S_0$ is the effective surface recombination velocity at $x = 0$, $S_d$ is the effective surface recombination velocity at $x = d$, and $S_{SC}$ is the recombination velocity in the SCR on the back surface.

In order to find the distribution of the excess charge carriers with the coordinate $x$, which is perpendicular to the surface, we will look for a solution of the diffusion equation in the form

$$\Delta n(x) = C_1 \exp\left(-\frac{x}{L}\right) + C_2 \exp\left(\frac{x}{L}\right) . \tag{8}$$

This Ansatz is a reasonable approximation for a solar cell where most of the incident light intensity is absorbed within a thin layer near the front surface. To determine the coefficients $C_1$ and $C_2$, we will use the following boundary condition on the front surface

$$\Delta n(x=0) = C_1 + C_2 . \tag{9}$$

and on the back surface:

$$S_{SC}\left(C_1 e^{-d/L} + C_2 e^{d/L}\right) = \frac{D}{L}\left(C_1 e^{-d/L} - C_2 e^{d/L}\right). \tag{10}$$

In general case,

$$C_2 = C_1 \frac{(D/L - S_{SC})e^{-d/L}}{(D/L + S_{SC})e^{d/L}} . \tag{11}$$

Using (9) and (11) we obtain

$$\Delta n(x=0) \cong C_1\left(1 + \frac{D/L - S_{SC}}{D/L + S_{SC}} \exp\left(-\frac{2d}{L}\right)\right) . \tag{12}$$

In a similar manner, we obtain a relation between $\Delta n(x = d)$ and $C_1$

$$\Delta n(x=d) = C_1 \frac{D}{L} \cdot \frac{\left(1 + e^{-d/L}\right) + S_{SC}\left(1 - e^{-d/L}\right)}{(D/L + S_{SC})e^{d/L}} . \tag{13}$$

In this paper, we will limit ourselves to the case $\Delta n$ = const. This approximation also works when the rates of bulk and surface recombination are significantly lower than the rate of recombination in the SCR, because then their contribution can be neglected. Taking into account expressions (12) and (13) in (7), we get

$$J_{SC} = q\left[\frac{d}{\tau_{eff}^b(x=0)} + S_0 + b_d S_{SC} + S_d\right]\Delta n(x=0) , \tag{14}$$

where the parameter $b_d$ is found from the equation

$$b_d = \frac{2(D/L)e^{-d/L}}{D/L + b_d S_{SC}(\Delta n) + (D/L - b_d S_{SC}(\Delta n))e^{-2d/L}} , \tag{15}$$

and $S_{SC}(\Delta n)$ is determined by (2).

As can be seen from (14), in this case it is possible to introduce an effective SCR recombination velocity

$$S_{SCd} = b_d S_{SC}(\Delta n) . \tag{16}$$

Similarly, we can consider the case when the SCR exists on the surface $x = 0$. In this case, the recombination-generation balance equation has the form

$$J_{SC} = q \left[ \frac{d}{\tau_{eff}^b(x=d)} + S_0 + b_0 S_{SC} + S_d \right] \Delta n(x=d) , \qquad (17)$$

where the coefficient $b_0$ is given by

$$b_0 = 2 \frac{D}{L} \cdot \frac{e^{d/L}}{D/L - b_0 S_{SC}(\Delta n) + (D/L + b_0 S_{SC}(\Delta n)) e^{2d/L}} . \qquad (18)$$

In the same way as before, it is possible to introduce the effective SCR recombination velocity on the surface $x = 0$:

$$S_{SC0} = b_0 S_{SC}(\Delta n) \qquad (19)$$

To demonstrate the dependence of the excess concentration on the $x$-coordinate, let us consider a simpler case where the recombination velocity on the back surface is determined by the surface recombination. In this case, the dependence excess carrier concentration is

$$\Delta n(x) = \Delta n(x=0) \left( e^{-x/L} + \frac{D/L - S_{deff}}{D/L + S_{deff}} e^{x/L} \right), \qquad (19)$$

where $S_{deff} = b_{ds} S_d$, and

$$b_{ds} = 2 \frac{D}{L} \cdot \frac{e^{-d/L}}{D/L + S + (D/L - S) e^{-2d/L}} . \qquad (20)$$

Shown in Fig. 1(a) is the dependence of the excess carrier concentration in the base on the $x$ coordinate, obtained from (19) using the following parameters: $\tau_{SRH} = 10^{-2}$ s, $d = 1.5 \cdot 10^{-2}$ cm, $L = 0.33$ cm of the back surface. The curves are parameterized by the back surface recombination velocity, which for the case of curves 1-6, respectively, was assumed equal to 1, 10, $10^2$, $10^3$, $10^4$, and $10^5$ cm/s.

As can be seen from Fig. 1(a), the excess concentration of minority carriers in the base changes linearly with $x$. At $S_d \leq 10$ cm/s, it changes weakly, in particular, for $S_d = 10$ cm/s, $\Delta n$ decreases by one percent. In the case when $S_d = 10^4$ cm/s it decreases by more than an order of magnitude, and when $S_d = 10^5$ cm/s it varies by approximately two orders of magnitude.

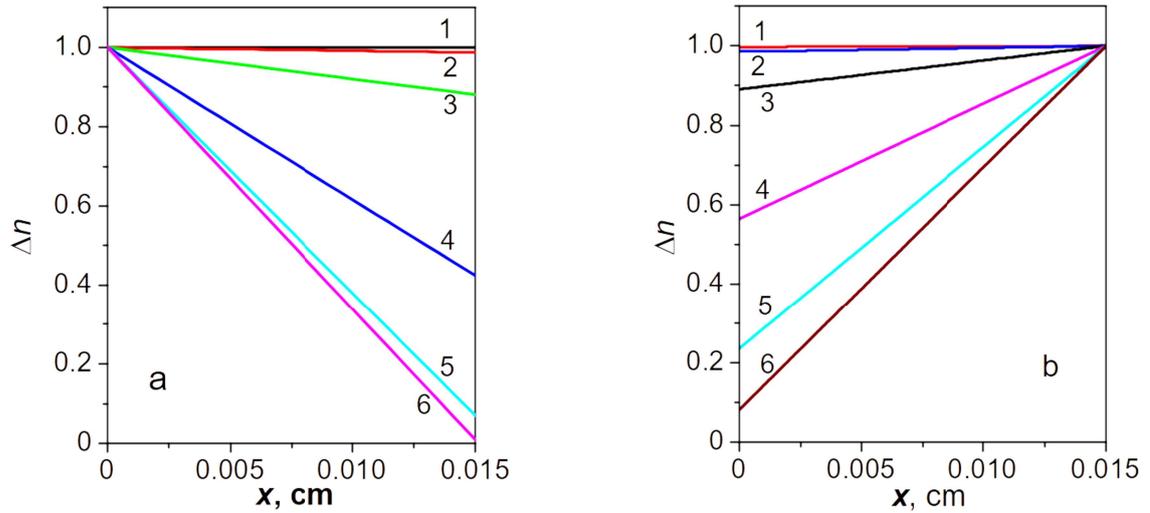

Fig 1. Excess carrier concentration as a function of coordinate $x$ in the case when the surface recombination velocity on (a) the back surface and (b) the front surface is set to 1(1), 10(2), $10^2$(3), $10^3$(4), $10^4$(5), and $10^5$ cm/s (6).

Fig. 1(b) shows the dependence $\Delta n(x)$ for the case when the surface recombination rate $S_0$ is high on the frontal surface $x = 0$. As can be seen from the figure, in this case the value of $\Delta n(x)$ decreases according to a linear law when approaching the surface at $x = 0$. Although the dependencies shown in Figs. 1(a) and (b) are symmetrical, Fig. 1(b) is interesting in that, despite the fact that the light is absorbed by the semiconductor from the side of the front surface, the concentration of excess pairs decreases when approaching it. This dependence is completely controlled by the value of the surface recombination velocity on this surface.

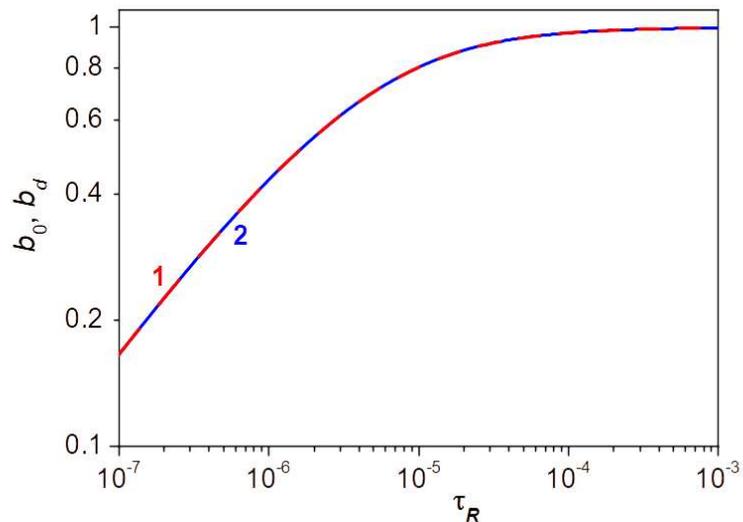

Fig. 2 The coefficients $b_0$ (1) and $b_d$ (2), which describe a decrease of the effective SCR recombination velocity, on the charge carrier lifetime in the base region.

Fig. 2 shows the dependence of the coefficients $b_0$ and $b_d$ on the value of $\tau_R$. As can be seen from the figure, the two curves coincide and decrease by almost an order of magnitude as the lifetime decreases within the specified range. Note that the SCR recombination is active on the rear surface in silicon SCs based on the structures with rear metallization (see [16]), while in structures with a conventional geometry it is operative on the front surface (see work [13]).

The magnitude of the SCR recombination velocity is primarily influenced by the value of the SCR lifetime $\tau_R$, the ratio of hole and electron capture cross sections $b_r$ and the doping level $n_0$. It increases with decreasing values of $\tau_R$ and $b_r$ and increases with increasing $n_0$. The value of SCR lifetimes $\tau_R$ for different rectifying structures varies widely, usually from $10^{-4}$ to $10^{-7}$ s, and is three to four orders of magnitude smaller than the Shockley-Reed-Hall lifetime in the neutral volume. We will not discuss the reason for this difference now, but will accept it as an experimental fact [13-15].

Fig. 3 shows the calculated theoretical dependences of the recombination velocity in the SCR on the excess concentration of electron-hole pairs $\Delta n$ for a SC with carrier lifetime in the SCR which differs by an order of magnitude ($10^{-6}$ s (a) and $10^{-7}$ s (b)). Curves 1 and 2 illustrate the dependence of $S_{SC0}$ and $S_{SCd}$ on $\Delta n$, respectively. Curve 3 represents the dependence of $S_{SC}(\Delta n)$, and curve 4 in Fig. 3a is an approximation to $S_{SC}(\Delta n)$ in the form of $S_a \sim \Delta n^{-0.6}$.

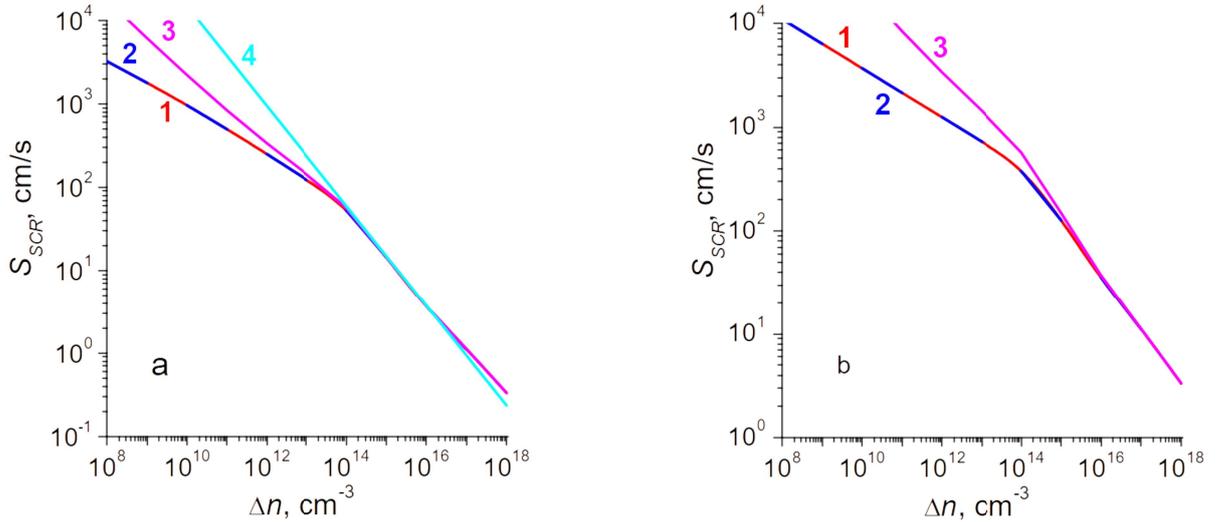

Fig. 3. The effective SCR recombination velocity on the excitation level for the SCR lifetime set to (a) $10^{-6}$ s and (b) $10^{-7}$ s. The curves 1 and 2 describe $S_{SC0}$ та $S_{SCd}$ vs. $\Delta n$, the curve 3 shows $S_{SC}(\Delta n)$, and the curve 4 in panel (a) is an approximation of the form $S_a \sim \Delta n^{-0.6}$.

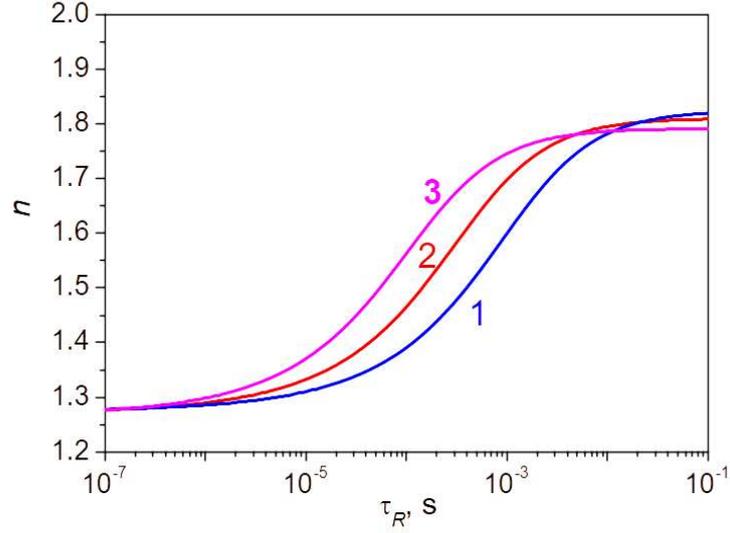

Fig. 4. The ideality factor associated with the SCR recombination as a function of the lifetime of charge carriers in the base. The curves differ in the ratio of hole and electron capture cross sections: $b_r$=0.1, 1, and 10, respectively, for blue (1), red (2), and crimson (3) curves.

Fig. 4 shows the ideality factor related to the SCR recombination

$$n = \frac{q}{kT}\left(\frac{d\ln(J)}{dV}\right)^{-1}, \qquad (21)$$

$$J = qS_{SCd}(\Delta n)\Delta n, \quad \Delta n(V) = -\frac{n_0}{2} + \sqrt{\left(\frac{n_0}{2}\right)^2 + n_i(T)^2\left(e^{qV/kT}-1\right)}, \qquad (22)$$

vs. the SCR lifetime in silicon barrier structures. The parameter of the curves is $b_r$, i.e. the ratio of the hole-to-electron capture cross sections. Figure 4 shows that the value of $n$ is always less than 2; its minimum value is close to 1.3. The reduction of the ideality factor begins at a significantly longer SCR lifetimes than the reduction of the coefficients $b_d$ or $b_0$.

It can also be seen from Fig. 4 that the larger the value of the hole capture coefficient compared to the electron capture coefficient, the greater the values of the ideality factor for the SCR.

As numerical estimates show, at $V = 0.1$ V, for typical parameters of silicon barrier structures with long lifetimes, the value of $J$ is about $10^{-8}$ A/cm², while the current density, which is determined by the shunt resistance at its value of $3\cdot10^4$ Ohm·cm², equals $3\cdot10^{-6}$ A/cm². In order for the contribution from the SCR recombination current density to dominate, the value of the shunt resistance should be at least three orders of magnitude larger. Since the following theoretical approach will be applied to the structures mentioned above, and in them the shunt resistance values lie in the range from $10^3$ to $10^5$ $\Omega \cdot cm^2$, this circumstance must be taken into account.

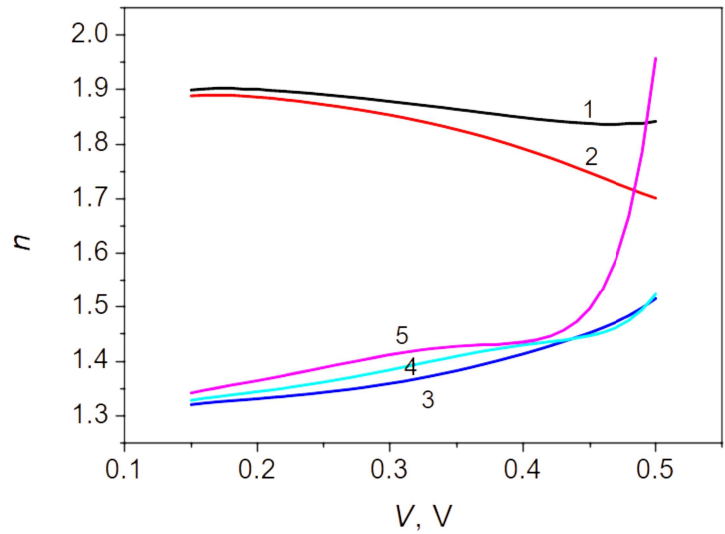

Fig. 5. The ideality factor of a *p-n* junction diode vs. applied voltage in a silicon structure with short (curves 1 and 2) and long (curves 3-5) diffusion length. The parameters used to build curves 1 and 2 are: $d_1 = 350$ μm, $n_{01} = 3 \cdot 10^{15}$ cm$^{-3}$, $L_1 = 30$ μm, $\tau_{R1} = 1.2 \cdot 10^{-8}$ s, (curve 1), $d_2 = 380$ μm, $n_{02} = 3.1 \cdot 10^{15}$ cm$^{-3}$, $L_2 = 166$ μm, $\tau_{R2} = 1.2 \cdot 10^{-6}$ s (curve 2), $T = 300$ K. The curves 3-5 are built with the parameter values $d = 0.015$ cm, $L = 0.33$ cm, $n_0 = 10^{15}$ cm$^{-3}$, $T = 300$ K, $N = 10^{12}$ cm$^{-2}$, $\tau_R = 5 \cdot 10^{-7}$ s (3), $5 \cdot 10^{-6}$ s (4) and $5 \cdot 10^{-5}$ s (5), respectively.

We described the curves in Figs. 4 in such detail to show that the ideality factor due to SCR recombination in this case depends on several parameters and variables, in particular, the SCR life time, the ratio of the hole and electron capture cross sections, the excess concentration of electron-hole pairs, etc. Therefore, its value can be given only for illustrative purposes. As for the effect of SCR recombination on the characteristics of silicon structures, it should be investigated by analyzing, first of all, the dependence of the SCR recombination velocity on the parameters mentioned above.

In simulation programs of a sufficiently high level, the ideality factor is not used, but the problem is solved based on general equations, so the question of the ideality factor value does not arise (see, for example, the program PC1d and its subsequent versions).

Unless stated otherwise, the following parameters will be used in the numerical evaluations and in the construction of the curves: $\tau_{SRH} = 10^{-2}$ s, $d = 150$ μm, $n_0 = 10^{15}$ cm$^{-3}$, $T = 300$ K.

At the end, we formulate the criteria for the validity of the results presented in the paper [1]. The first condition is the requirement that the thickness of the base region $d$ significantly exceeds the diffusion length $L$. In this case, only one surface is involved and $\Delta n = \Delta n(x = 0)$. The second condition is the requirement that the excess concentration of electron-hole pairs $\Delta n$ be much smaller than the equilibrium concentration of the majority charge carriers $n_0$. In this case, the integral function in expression (1) is not significantly deformed due to $\Delta n$ and the value of the non-ideality coefficient is close to 2. In most cases, the mentioned criteria are fulfilled for barrier structures based on all semiconductors except modern silicon. But in Ref. [14] we considered samples of silicon SCs for which these criteria are fulfilled. Fig. 5 shows the ideality factor as a function of the applied voltage for two SCs, one with an n-type base and the other with a p-type base, in which the diffusion lengths are much smaller than the thickness of the base (diffusion lengths are 166 and 50 μm, respectively, and base thickness equals 380 and 350 μm). In Fig. 5(b), theoretical curves are shown for the ideality factor of SCR recombination on the applied voltage for a high-efficiency silicon SC sample, in which the diffusion length is an order of magnitude greater than the base thickness (respectively 1800 and 150 μm). The parameter of the curves is the $b_r$ value. As can be seen from the given figure, in this case, for SCs with small diffusion lengths, there is a weak dependence on the applied voltage, the values of the ideality factor are close to 2 and do not decrease significantly with increasing voltage. At the same time, the values of the ideality factor for SCs with large diffusion length behave in accordance with the theory outlined above, increasing from small values of the order of 1.3 to values greater than 1.4 at $V = 0.4$ V. If for the former, the values of the ideality factor decrease by 3% and 5% in in the specified range of voltage, then for the latter, they increase by 8% on average.

As compared to other semiconductors, monocrystalline silicon has very long Shockley-Reed-Hall lifetimes, which can reach values of the order of 100 ms. Diodes with a thin base and p-i-n structures are routinely produced based on this material. Although monocrystalline silicon is an exception in this regard, the majority of solar cells are based on it, devices with silicon p-n junctions are widely used in semiconductor microelectronics, and the effects considered in this paper are important for their operation.

### 3. Experimental determination of $S_{SCd}(\Delta n)$ in silicon barrier structures

In this section, we will describe the theoretical method of experimentally finding the surface recombination velocity $S_{SCd}(\Delta n)$ as a function of the excess concentration and apply it in practice. For its implementation, it is necessary to subtract the contribution associated with all other recombination mechanisms from the total effective lifetime of charge carriers. The contribution from radiative recombination and interband Auger recombination is known and does not change from sample to sample, while the contribution from Shockley-Reed-Hall recombination and from surface recombination must be determined using known methods of their investigation, or by varying their values, fit experimental and theoretical dependences of $J_L(V_{OC})$. For a complete fit, it is necessary to pre-set the values of $\tau_R$ and $b_R$, which allows to find approximately the values of $\tau_{SRH}$ and $S$. When performing this procedure, it is expedient to set the initial value of $b_R$ to 1.

This was done first using the experimental dependences for the short-circuit current density on the applied voltage $J_L(V_{OC})$, described in our work [16]. The experimental dependence of $S_{SC}(\Delta n)$ determined in this way is shown in Fig. 6 (symbols).

To compare the obtained experimental dependence with the theory, two circumstances should be taken into account. First, in cases where the value $\tau_R$ is small enough, in the region of sufficiently large values of $\Delta n$, in addition to SCR recombination, it is necessary to take into account the bulk recombination in that part of the SCR that has become neutral. Secondly, the model with $\tau_R$ = const is not realistic at medium and large excess concentrations. Much better is the model that describes the dependence of $\tau_R^{-1}$ on the $x$-coordinate in the SCR by a Gaussian

$$\tau_R^{-1}(x) = \tau_{Rm}^{-1} \exp\left(-\frac{(x-x_m)^2}{2\sigma^2}\right), \tag{23}$$

in which $x_m$ is the position of the maximum, $\sigma$ is the variance, $\tau_{Rm}$ is the lifetime at the ~~point of Gaussian~~ maximum, SCR recombination velocity becomes

$$S_{eff}^{sc} = \int_0^{d_{eff}} \frac{\tau_{Rm}^{-1}\exp\left(-\frac{(x-x_m)^2}{2\sigma^2}\right)(n_0+\Delta n)dx}{(n_0+\Delta n)e^{y(x)} + b_r(p_0+\Delta n)e^{-y(x)}}. \tag{24}$$

The effective thickness $d_{eff}$ is found from

$$\tau_{Rm}^{-1}\exp\left(-\frac{(d_{eff}-x_m)^2}{2\sigma^2}\right) = \tau_{SRH}^{-1}, \tag{25}$$

It is the effective SCR thickness up to which integration should be performed to correctly calculate $S_{eff}^{sc}$. The reason is that the experimental SCR lifetimes are lower than the SRH lifetime by a few orders of magnitude. Therefore, increasing the upper integration limit even by several times has very little effect on the SCR recombination velocity value obtained.

The next problem is finding the SCR recombination velocity taking into account the excess concentration gradient $\Delta n$. Previously, we used an expression

$$b_d = \frac{1}{\cosh\left(\dfrac{d}{L}\right) + S_{SC}(\Delta n)\dfrac{L}{d}\sinh\left(\dfrac{d}{L}\right)}. \tag{26}$$

The value of $S_{SC}(\Delta n)$ in this equation is determined by the expression (2) and depends on the value of $y_l$, the dimensionless potential at the boundary of the SCR and the quasi-neutral region. Note that expression (2) allows to find the recombination velocity only in that part of the SCR where the magnitude of the non-dimensional potential is greater than or equal to 1. In contrast, expression (25) allows gives the total recombination velocity both in that part of the SCR where the electrostatic potential is greater than or equal to 1 and in that part where the band straightening took place. At sufficiently large excess concentrations, the second term may exceed the first. Taking this into account, expression (26) should be generalized by replacing the value of $S_{SC}(\Delta n)$ with $S_{sum}^{SC}$.

The parameter $b_{dsum}$ that describes the ratio of excess concentrations on the surfaces $x = 0$ and $x = d$ in this case is defined as

$$b_{dsum} = \frac{1}{\cosh\left(\dfrac{d}{L}\right) + S_{sum}^{SC}\dfrac{L}{d}\sinh\left(\dfrac{d}{L}\right)}. \tag{27}$$

The quantity $S_{effd}^{SC}$, which takes into account the gradient effect including the contribution from the region where the bands are straightened, is now found from

$$S_{effd}^{SC} = b_{dsum} S_{eff}^{SC}(\Delta n) \ . \tag{28}$$

Comparing the values (26) and (27) with each other, the following differences should be noted. First, due to the fact that $S_{effd}^{SC}$ is greater than $S_{SC}(\Delta n)$, when $x_m$ and $\sigma$ are large enough, the value of $b_{dsum}$ should be slightly less than $b_d$. However, in the case of sufficiently small $x_m$ and $\sigma$, the value of $S_{SC}(\Delta n)$ will be decreased relative to $S_{SC}(\Delta n)$, which will lead to an increase in $b_{dsum}$. As the calculations show, the second effect dominates ins silicon SCs described in [13, 16], i.e. the resulting value of $b_{dsum}$ increases slightly. However, this effect is relatively small and in the relevant range of not too big $\Delta n$ it is about 10%.

The theoretical values obtained using Eq. (28) must be fitted to the experimental curves (symbols in Figs. 6-8). Our calculations showed that for the case of SC described in [16], agreement between the experimental $S_{SC}(\Delta n)$ data and the theoretical equation (28) is achieved for $\tau_{Rm} = 8.4\cdot 10^{-6}$ s, $b_r = 0.1$, $x_m = 1.8\cdot 10^{-5}$ cm, $\sigma = 1.5\cdot 10^{-5}$ cm, and $d_{eff} = 8.3\cdot 10^{-5}$ cm (see Fig. 6).

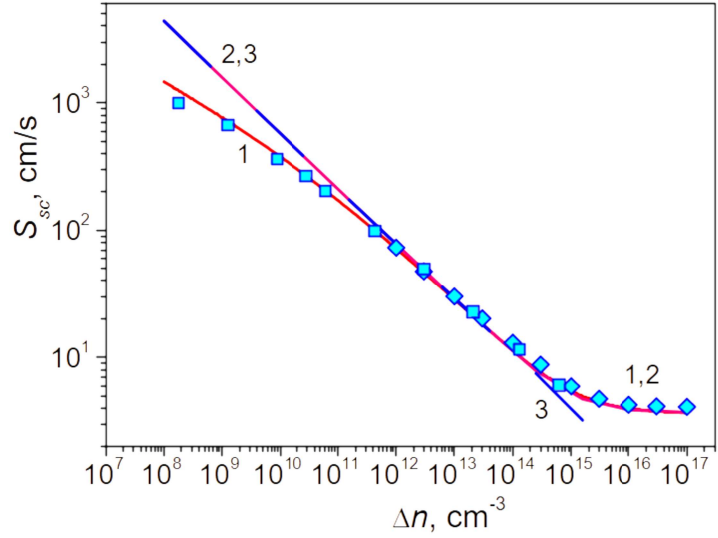

Fig. 6. Experimental dependence of $S_{SC}(\Delta n)$ (symbols). Theoretical curves 1 and 2 for the case when $\tau_{Rm} = 8.3\cdot 10^{-6}$ s, $b_r = 0.1$, $x_m = 1.8\cdot 10^{-5}$ cm, $\sigma = 1.5\cdot 10^{-5}$ cm, $d_{eff} = 8.35\cdot 10^{-5}$ cm are obtained using the data given in [16]. Curve 3 describes the SCR recombination velocity without taking into account the contribution of the initial part of the SCR that has become neutral.

The curves $S_{effd}^{SC}$ vs. $\Delta n$ built in the interval from $10^8$ to $10^{17}$ cm$^{-3}$ using Eq. (28) and (24) are represented by curves 1 and 2, respectively. Curve 3 shows the SCR recombination velocity given by an expression

$$S_w^{sc}(\Delta n) = \int_0^{w(\Delta n)} \frac{\tau_{Rm}^{-1} \exp\left(-\frac{(x-x_m)^2}{2\sigma^2}\right)(n_0 + \Delta n)dx}{(n_0 + \Delta n)e^{y(x)} + b_r(p_0 + \Delta n)e^{-y(x)}}, \qquad (29)$$

which does not account for the contribution of that initial part of the SCR that became neutral. As can be seen from the comparison of curves 2 and 3, at $\Delta n > 10^{14}$ cm$^{-3}$ the curves diverge, and at $\Delta n = 10^{16}$ cm$^{-3}$ the value of $S_{eff}^{SC}$ is equal to 3.9 cm/s, while the value of $S_w^{SC}$ is 1.14 cm/s. That is, their difference, which describes the contribution of that SCR part that has become neutral, is equal to 2.76 cm/s. At $\Delta n \geq 10^{15}$ cm$^{-3}$, the function $S_{eff}^{SC}(\Delta n)$ tends to saturate, while $S_w^{SC}(\Delta n)$ continues to decrease with a slope close to 0.5.

Fig. 7 shows the experimental (symbols) and theoretical (curves) results for the SCR recombination velocity, obtained for the materials described in [13]. Curves 1 and 2 are obtained using Eq. (28) and (24), respectively. Agreement between experiment and theory was achieved using the following parameters: $\tau_{Rm} = 2.5 \cdot 10^{-7}$ s, $b_r = 50$, $x_m = 3 \cdot 10^{-5}$ cm, $\sigma = 5 \cdot 10^{-5}$ cm, $d_{eff} = 2.47 \cdot 10^{-4}$ cm. The curves 2 and 3 exhibit similar trends as in the case analysed in Fig.6.

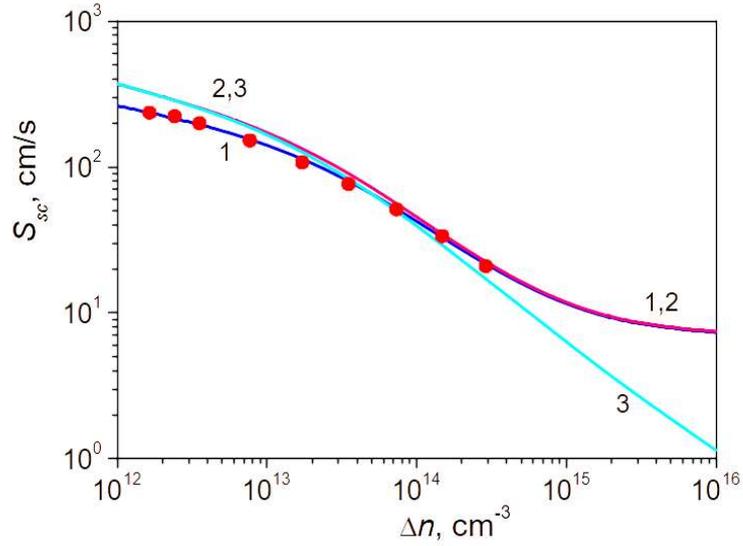

Fig. 7. SCR recombination velocity $S_{SC}(\Delta n)$, obtained using experimental data of [13]. Symbols are experimental data. Curves 1 and 2 are, respectively, the theoretical $S_{eff}^{sc}$ and $S_{effd}^{sc}$ plots. The curve 3 represents SCR recombination velocity obtained without accounting for that part of the SCR that has become neutral. The theoretical curves are obtained with the following parameter values: $\tau_{Rm} = 2.5 \cdot 10^{-7}$ s, $b_r = 50$, $x_m = 3 \cdot 10^{-5}$ cm, $\sigma = 5 \cdot 10^{-5}$ cm, $d_{eff} = 2.47 \cdot 10^{-4}$ cm.

Finally, let us use the procedure described above to find the experimental value of $S_{SC}$ and its dependence on the excess concentration of charge carriers for the SC with record photoconversion efficiency described in the work of Yoshikawa et al. [17]. In our work [22], we have already theoretically modeled the key characteristics of this SC, taking into account recombination in the SCR. In order to perform the procedure described above for finding the experimental value of the recombination rate in the SCR, we used Fig. 4b from [17] and the following parameters obtained by fitting the theoretical dependence (23) in [22]: $\tau_{Rm} = 1.4 \cdot 10^{-5}$ s, $b_r = 0.1$, $x_m = 2.5 \cdot 10^{-5}$ cm, $\sigma = 4.5 \cdot 10^{-6}$ cm, $d_{eff} = 4.18 \cdot 10^{-5}$ cm.

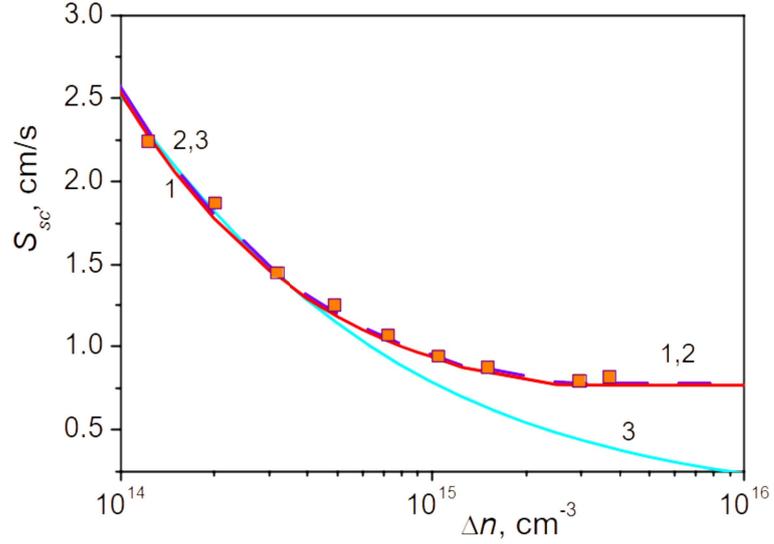

Fig. 8. Symbols: experimental SCR recombination velocity $S_{exp}(\Delta n)$. The theoretical curves 1 and 2 are obtained using the data given in the work [17] based on the formulas (28) and (24) with $\tau_{Rm}$ =1.4·10$^{-5}$ s, $b_r$ = 0.1, $x_m$ = 2.5·10$^{-5}$ cm, $\sigma$ = 4.5·10$^{-6}$ cm, $d_{eff}$ =4.18·10$^{-5}$ cm. The curve 3 describes the SCR recombination velocity without accounting for that initial part of the SCR that has become neutral.

The curves 2 and 3 behave similar to the case from Fig. 6. Fig. 8 shows the so obtained experimental and theoretical relations obtained using formulas (28) and (24), curves 1 and 2, respectively. As can be seen from the figure, at $\Delta n > 10^{14}$ cm$^{-3}$, the experimental and theoretical dependences of $S_{SC}$ constructed in different approximations, completely coincide. This is due to the large values of $\tau_{Rm}$ in this case and the lower-bounded experimental value $\Delta n$ of the order of $10^{14}$ cm$^{-3}$, which was realized in [17].

The general conclusion that can be drawn by analyzing Figs. 6-8 is that within the excess concentration range $\Delta n > 10^{14}$ cm$^{-3}$, the contribution of bulk recombination in that part of the SCR which has become neutral, exceeds the contribution of recombination in the SCR itself and must be taken into account in the calculations of $S_{eff}^{sc}$. The obtained results also indicate that the Gaussian distribution of the inverse SCR lifetime satisfactorily describes the experimental distribution of deep levels in the SCR of these SCs, as evidenced by the good agreement between theoretical curves and experimental data.

It should be noted that despite the fact that the value of the effective recombination velocity in the SCR in this case is small, due to its very slow decline in the vicinity of the maximal photoconversion efficiency, it can play a significant role along with other recombination processes and affect the value of the photoconversion efficiency. This will be discussed in more detail in the next section.

**4. The effect of the Δ*n*(*x*) dependences associated with a high SCR recombination velocity on the effective time of excess charge carriers in the base of the barrier structure**

The effective lifetime of excess charge carriers is an important characteristic of the quality of silicon barrier structures and its research has been quite intensive among scientists who are engaged in the development of silicon SCs and are looking for ways to increase their efficiency. There are several approaches to its determination, and one of them, as shown in a number of works, consists in studying the dependence of the short-circuit current on the open circuit voltage [16,17]. From a physical point of view, the study of $J_L(V_{OC})$ dependences is similar to the study of dark *I-V* curves and provides information about the SCR recombination. Their only difference is that $J_L(V_{OC})$ curves, in contrast to dark *I-V* curves, do not contain information about series resistance.

Unfortunately, today there are few works, with the exception of [16,18], in which studies of $J_L(V_{OC})$ dependences were performed in a wide range of illumination intensities, starting from values when $V_{OC}$ is small. In the work of Cuevas and Kerr [18], the interpretation of the obtained results was carried out using a simplified approach, which was based on the use of biexponential *I-V* curves, one of them was considered to be the contribution from SCR recombination with an assumed ideality factor of 2. In our work [16], the interpretation of the results of the study of the SunPover SCs with reverse metallization was performed in a more general form. In work [17], these dependencies, or rather, the dependencies of illumination on the open circuit voltage, were measured starting from $V_{OC}$ values equal to 0.53 V.

The short-circuit current of a SC with a unit area can be written as [16]

$$I_L = \frac{qd}{\tau_{eff}(\Delta n_{OC})} \Delta n_{OC} + \frac{V_{OC}}{R_{sh}} , \qquad (30)$$

where

$$\Delta n_{OC} = -\frac{n_0}{2} + \sqrt{\frac{n_0^2}{4} + n_{i0}^2 e^{\Delta E_g / kT} \left(e^{qV_{OC}/kT} - 1\right)} . \qquad (31)$$

Here $n_{i0}$ is the intrinsic concentration in the absence of the bandgap narrowing effect, and $\Delta E_g$ is its value [19]. The dependence of the dark current on the applied voltage is similar to the expression (30):

$$I_D(V) = \frac{qd\Delta n}{\tau_{eff}(\Delta n)} + \frac{V - IR_s}{R_{SH}} \quad . \tag{32}$$

The only difference between the two expression is that $I^P(V)$ depends on the series resistance $R^s$.

It follows from (32) that

$$\tau_{eff} = \frac{J_L(V_{OC}) - \frac{V_{OC}}{R_{sh}}}{qd\Delta n_{OC}} \quad . \tag{33}$$

For further calculations, we will need a theoretical expression for the effective lifetime in silicon $\tau_{eff}(\Delta n)$. The net lifetime is formed by intrinsic and extrinsic recombination mechanisms,

$$\tau_{eff}^{-1} = \tau_{intr}^{-1} + \tau_{extr}^{-1} \quad , \tag{34}$$

where the intrinsic lifetime $\tau_{intr}$ is formed by the radiative and Auger band-to-band recombination (see [17,20]), and the extrinsic lifetime $\tau_{extr}$ is formed, by SRH recombination, the non-radiative exciton Auger recombination assisted by deep impurities, surface recombination, and SCR recombination. Denoting the lifetimes associated with these processes, respectively, as $\tau_{SRH}$, $\tau_{eff}^b$, $\tau_{eff}^s$, and $\tau_{scr}$, we write

$$\tau_{extr} = \left( \left(\tau_{eff}^b\right)^{-1} + \tau_{SRH}^{-1} + \left(\tau_{eff}^s\right)^{-1} + \left(\tau_{scr}\right)^{-1} \right)^{-1} . \tag{35}$$

The effective bulk lifetime of charge carriers in the base region $\tau_{eff}^v$, discussed earlier, is defined as

$$\tau_{eff}^v = \left(\tau_{eff}^b\right)^{-1} + \tau_{SRH}^{-1} + \tau_{intr}^{-1} . \tag{36}$$

The value of the Shockley-Reed-Hall lifetime depending on the doping level and the excitation level in an n-type semiconductor is described by the expression

$$\tau_{SRH} \cong \frac{\tau_{p0}(n_0 + n_1 + \Delta n) + \tau_{n0}(p_1 + \Delta n)}{(n_0 + \Delta n)} , \tag{37}$$

where $\tau_{p0} = (V_p \sigma_p N_t)^{-1}$, $\tau_{n0} = (V_n \sigma_n N_t)^{-1}$, $V_p$ and $V_n$ are the mean thermal velocities of holes and electrons, $\sigma_p$ and $\sigma_n$ are their capture cross-section by the recombination centers of concentration $N_t$, and $n_1$, $p_1$ are the electron and hole concentrations when the Fermi energy coincides with the energy of the recombination center. Depending on the excess concentration of electron-hole pairs, the lifetime $\tau_{SRH}$ changes between the low-injection and high-injection extreme values.

The non-radiative exciton Auger recombination lifetime [21] is

$$\tau_{eff}^b = \tau_{SRH} \frac{n_x}{n_0 + \Delta n} \quad , \tag{38}$$

where $n_x = 8.2 \cdot 10^{15}$ cm$^{-3}$. In the appendix, the history of the appearance of this lifetime is considered in detail, the necessity of its use is substantiated, its place among other times is established, and its manifestation in silicon barrier structures with long lifetimes is illustrated on a concrete example.

For the Auger interband recombination lifetime, we used the empirical expression given in [20].

The surface recombination lifetime is

$$\tau_{eff}^s = d / S_s \quad , \tag{39}$$

where $S_s$ is the total recombination velocity on the front and the back surfaces.

Next, we specify the dependence of the surface recombination velocity on the level of excitation and on the level of doping. We will assume that the value is determined by the expression

$$S_s = S_{0s} \left( \frac{n_0}{n_p} \right)^m \left( 1 + \frac{\Delta n}{n_0} \right)^r \quad , \tag{40}$$

where $S_{0s}$ is the total value of surface recombination velocity on the front and back surfaces at a low level of excitation, $n_p$ is the initial value of the doping level, $m \approx 1$, and the value of $r$ for most silicon samples is also equal to 1. The lifetime due to the SCR recombination is defined as

$$\tau_{SCR} = \left( \frac{S_{SCd}}{d} \right)^{-1} \quad . \tag{41}$$

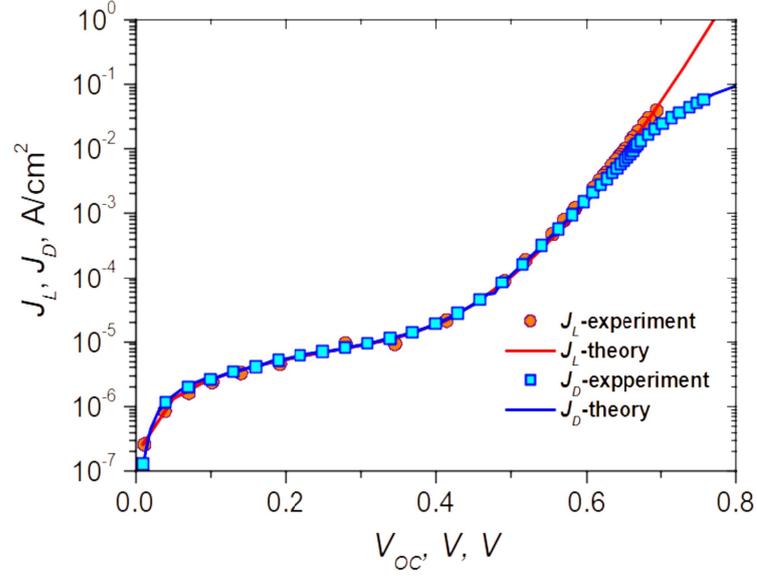

Fig.9. The dark current density vs. applied voltage and the short circuit current density vs. the open circuit voltage, as obtained in [13]. The symbols are the experimental data, the lines are the theoretical curves.

Now we will return to the interpretation of the results of the $J_L(V_{OC})$ measurements, taking into account the results of the first section of this paper, that is, using the $S_{SCd}(\Delta n)$ dependences. Fig. 6 shows the experimental curves $J_L(V_{OC})$, combined with the $J_D(V)$ curves for one of the SCs studied in [16]. As can be seen from the figure, they coincide up to $V_{OC}(V)$ values of 0.65 V, and diverge at higher values. They were calculated theoretically within the approach used in this work, when expression (2) was used once for the dependences of the recombination rate in the SCR, and the second time the expression $S_{SCd}(\Delta n)$ was applied.

Visually, there are no differences in the $J_L(V_{OC})$ and $J_D(V)$ curves calculated in the first and the second approximation. However, the theoretical dependences of $\tau_{eff}(\Delta n)$, obtained using the $J_L(V_{OC})$ curves, are shown in Fig. 7. At the values of $\Delta n$ less than $10^{14}$ cm$^{-3}$, they differ. Their difference is the bigger the smaller the value of $\Delta n$. The obtained result is consistent with the theory developed in the first chapter. The $\tau_{eff}(\Delta n)$ values calculated using the $S_{SCd}(\Delta n)$ dependences are larger.

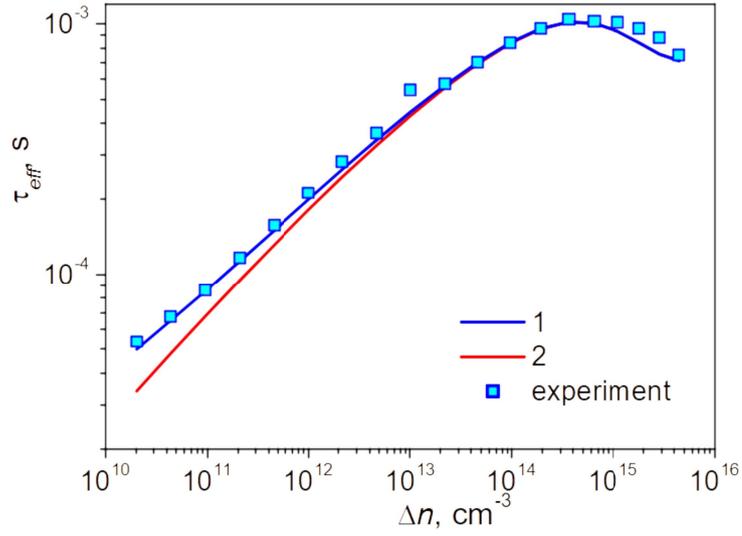

Fig. 10. Effective life time in silicon vs. the excitation level obtained theoretically with and without taking into account the change in the SCR recombination velocity in the presence of a concentration gradient (curves 1 and 2) and the experimental data built based on the values of $\tau_R$ and $b_r$ from the work [13].

Fig. 10 also shows the experimental $\tau_{eff}(\Delta n)$ curves obtained using the experimental values of $J_L(V_{OC})$. As can be seen from the figure, it is in good agreement with the theoretical dependence constructed using $S_{SCd}(\Delta n)$ formula. A procedure similar to the one used above was used in processing the dependences of the dark current on the applied voltage, measured in our work [13]. Theoretical curves of the effective lifetimes were constructed with and without taking into account the dependences of the SCR recombination velocity $S_{SC}(\Delta n)$ on the calculated effect. They are shown in Fig. 11. It also shows the experimental dependence of $\tau_{eff}(\Delta n)$ using the experimental values of $J_D(V)$ measured in this work, at voltage values lower than 0.6 V, when the dark I-V characteristic coincides with the $J_L(V_{OC})$ dependence. As can be seen from the figure, at $\Delta n$ less than $10^{15}$ cm$^{-3}$, which corresponds to the voltage less than 0.6 V, it is consistent with the theoretical dependence that takes into account the calculated effect, i.e. the experimental values of $\tau_{eff}(\Delta n)$ are larger.

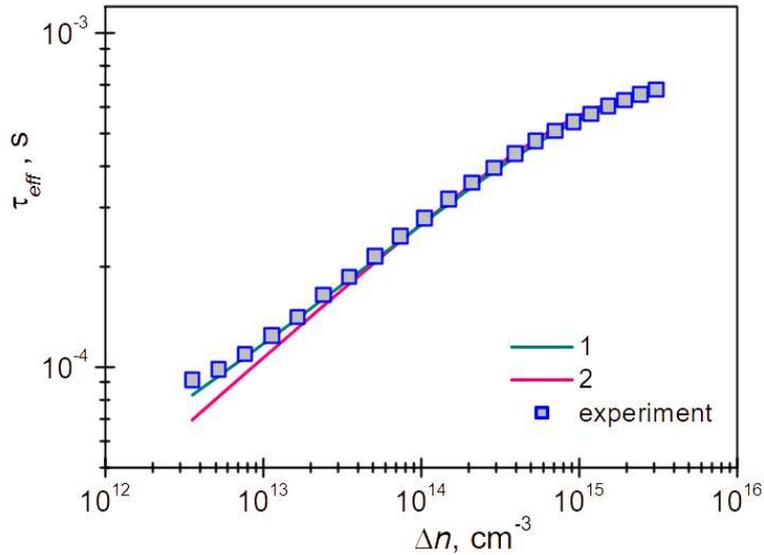

Fig. 11. Effective lifetime in Si SC base vs. the excitation level, as obtained with and without taking into account the variation of the SCR recombination velocity in the presence of the concentration gradient (curves 1 and 2) and the respective experimental curve obtained using the results and the values for $\tau_R$ та $b_r$ from [16].

## 5. The effect of SCR recombination on the characteristics of highly efficient silicon solar cells

Theoretical estimates, which are based on the assumption that the SCR recombination time in silicon solar cells are close to the SRH recombination time in the base, lead to the conclusion that the SCR recombination in silicon SCs should not affect their characteristics, unlike silicon diodes. This is due to the fact that the main characteristics of CE, such as photoconversion efficiency, open-circuit voltage, and others, are determined by their operation in the region of sufficiently large characteristic concentrations of excess electron-hole pairs under conditions of maximum extracted power and under conditions of open circles In this range of excess concentrations of charge carriers (of the order of $10^{15}$-$10^{16}$ cm$^{-3}$), according to estimates using the above assumption at $\tau_R = 10^{-3}$ s according to expression (2), the value of $S_{SC}(\Delta n)$ in the order of magnitude is equal to 0.01 cm/c, that is, it is significantly smaller than other recombination components.

However, the reality was not so optimistic. It was found that, as mentioned earlier, the recombination times in the SCR were significantly shorter, of the order of $10^{-4}$ to $10^{-7}$ s. The reasons for this are not always clear, especially in the case of HIT elements, in the manufacture of which high temperatures are not used. The estimation of the values of $S_{SC}(\Delta n)$ and $S_{SCw}(\Delta n)$ at $\tau_R = 10^{-7}$ s gives values of the order of 150 and 100 cm/c, which, as a rule, significantly exceeds the recombination contributions from other recombination mechanisms. Therefore, we will further calculate the effect of recombination in the SCR on the parameters of high-efficiency silicon SCs depending on the value of $\tau_R$. Furthermore, as can be seen from Figs. 6-8, in the range $\Delta n \geq 10^{15}$ cm$^{-3}$, the contribution of bulk recombination in that part of the SCR, which has become neutral, is added to the net SCR recombination velocity. Therefore, we will further calculate the effect of recombination in the SCR on the parameters of high-efficiency silicon SCs depending on the $\tau_R$ value, taking this into account. For this we will need the theoretical expressions for all recombination components in silicon given in the third chapter.

Using the theoretical approach from [16], the I-V relation in the presence of illumination,

$$I(V) = I_L - \frac{qd\,\Delta n}{\tau_{eff}(\Delta n)} + \frac{V + IR_s}{R_{SH}}, \qquad (42)$$

allows to find such fundamental characteristics of silicon SCs as photoconversion efficiency under AM1.5 conditions and open circuit voltage. We will obtain their dependence on the SCR recombination lifetime with the following parameters: $\tau_{SRH} = 10^{-2}$ s, $d = 150$ μm, $n_0 = 10^{15}$ cm$^{-3}$, $T = 298$ K, $R_S = 0.3$ Ω·cm$^2$, $R_{SH} = 3\cdot10^4$ Ohm·cm$^2$.

First, let's analyze the case when the value of $b_r \geq 1$. The calculation results are shown in Table 1. To produce these data, it was assumed that $b_r = 1$. The same table shows the values of the excess concentrations $\Delta n_m$ and $\Delta n_{OC}$, which are equal to the values of the excess concentration at the points of maximum power collection and in the open circuit regime.

Table 1. Key parameters of high-efficiency silicon SCs for different values of SCR lifetime

| $\tau_R$, s | $10^{-3}$ | $10^{-4}$ | $10^{-5}$ | $10^{-6}$ | $10^{-7}$ |
|---|---|---|---|---|---|
| $\eta$, % | 24.0 | 24.0 | 23.5 | 21.6 | 18.6 |
| $V_{OC}$, V | 0.705 | 0.705 | 0.704 | 0.692 | 0.626 |
| $\Delta n_m$, cm$^{-3}$ | 1.74·10$^{15}$ | 1.69·10$^{15}$ | 1.39·10$^{15}$ | 5.47·10$^{14}$ | 3.02·10$^{13}$ |
| $\Delta n_{OC}$, cm$^{-3}$ | 6.86·10$^{15}$ | 6.84·10$^{15}$ | 6.65·10$^{15}$ | 5.1·10$^{15}$ | 1.13·10$^{15}$ |

As can be seen from the table, the effect of SCR recombination on the photoconversion efficiency is greater than its effect on the open circuit voltage. And this is understandable, because the excess concentration in open-circuit regime exceeds the concentration of excess charge carriers under the condition of maximum extracted power. At the same time, the value of the recombination velocity in the SCR is smaller and therefore its influence on the open circuit voltage is smaller. The effect of SCR recombination on the photoconversion efficiency depending on the value of the SCR lifetime at $\tau_R = 10^{-3}$ s is absent, at $\tau_R = 10^{-4}$ s it is weak, at $\tau_R = 10^{-5}$ s it is moderate, at $\tau_R = 10^{-6}$ s it is strong, and at $\tau_R = 10^{-7}$ s it is extremely high.

It should also be noted that $\Delta n_m$ in this case is significantly decreased, which correlates with a decrease in the photoconversion efficiency due to an additional increase in the SCR recombination velocity.

In all cases, except when $\tau_R = 10^{-7}$ s, in the maximum power regime, the gradient effects considered in this paper are insignificant. In the case when $\tau_R = 10^{-7}$ s and $\Delta n_m = 4.7 \cdot 10^{13}$ cm$^{-3}$, the value of $S_{SC}$ is 794, and the value of $S_{SCd}$ is 480 cm/s, that is, the gradient effect is already fully manifested.

If we compare the values for $\tau_R$ obtained in section 4 for SCs described in works [13] and [16] with Table 1, it can be seen that they correspond to a strong and moderate effect of SCR recombination on photoconversion efficiency.

We will analyze the case when $b_r < 1$ using the example of work [17]. In this work, the value of photoconversion efficiency of 26.6% was obtained. The maximum power selection point in this case was at $\Delta n = 3 \cdot 10^{15}$ cm$^{-3}$. At this point, the SCR recombination velocity according to Fig. 11 is 0.8 cm/s, and the sum of all other velocities, as shown by the calculation using the parameters given in the previous section, is 2.63 cm/s. That is, they are of the same order of magnitude, taking into account this, the rate of recombination in the SCR in this case should affect the efficiency of photoconversion. Calculation of the photoconversion efficiency of this SC in the absence of recombination in the SCR gives a value of 26.9%, which is 1% higher than the achieved value. Thus, in the case when the ratio $b_r < 1$, the recombination in the SCR affects the photoconversion efficiency of highly efficient silicon SCs more strongly than when $b_r \geq 1$, even in the case of not too small values of $\tau_R$.

The paper also contains two appendices. In the first appendix, some issues related to SCR recombination are considered in more detail. In the second appendix, a detailed analysis of the problem related to the need to take into account, among the recombination mechanisms in silicon, the non-radiative excitonic Auger recombination mediated by deep impurities, is performed.

**Conclusions**

The main result of this work is that when fitting the experimental results in silicon barrier structures with long bulk lifetimes, it is necessary to take into account that at sufficiently small excitation levels, high values of the recombination velocity in the SCR lead to a spatial dependence of the distribution of the excess concentration of electron-hole pairs in the base regions. This, in turn, renormalizes the value of the SCR recombination velocity, leading to its decrease. Taking into account this effect, the ideality factor is no longer a constant close to 2, but becomes a function that depends on several parameters and varies within a rather broad range (from ca. 1.3 to 1.8) depending on the value of the SCR lifetime. In this work, the calculations were performed under the assumption that the Shockley-Reed-Hall lifetimes in the base regions are large enough and greater than a millisecond, which leads to constant values of the excess concentration in these regions at small values of surface and SCR recombination velocities.

The paper shows that the manifestation of this effect leads to an increase in the effective lifetime of excess charge carriers at low values of the excess concentration. Theoretical calculations were confirmed experimentally.

A method is proposed and implemented to obtain SCR recombination velocity from the experimental data on silicon barrier structures. A comparison with the theory was carried out, which showed that in the region of sufficiently low values of $\Delta n$, the limitation of the SCR recombination velocity is in agreement with the considered effect.

It is shown that due to the decrease in the SCR thickness with $\Delta n$, a part of the initial SCR becomes quasi-neutral, and in this part, bulk recombination occurs with the lifetime $\tau_R$. Due to the small SCR lifetimes ($\leq 1$ μs) in real silicon $p^+ - n - n^+$ structures, the value of this recombination velocity $S_r$ can exceed the recombination velocity $S_{SC}$ in the SCR, which is determined by expression (2). This happens at sufficiently large values of $\Delta n$. Therefore, in the general case, the effective SCR recombination velocity is determined by their sum, i.e. $S_{eff}^{SC} = S_{SC} + S_r$.

Specific assessments of the impact of SCR recombination on the key characteristics of high-efficiency silicon solar cells, in particular, on the photoconversion efficiency and open circuit voltage, are made, and it is shown that this effect must be taken into account if the values of the lifetime in the SCR are of the order of less than $10^{-5}$ s; the lifetimes of this order of magnitude and even shorter are typical for silicon SCs.

It is shown that the cases when $b_r \geq 1$ and $b_r < 1$ should be considered separately. In the latter case, the effect of SCR recombination on the efficiency of photoconversion is stronger than in the former case.

In addition to the questions above, a detailed analysis of the problem related to the need to take into account the non-radiative exciton Auger recombination with the participation of deep impurities is performed in the appendix. First, its origin and its place alongside other recombination mechanisms are described retrospectively, and then a comparison of the theoretical dependencies for the effective lifetime in the silicon bulk with the experiment is made. It confirms the importance of the non-radiative exciton recombination mechanism.

**Appendix 1**

The electroneutrality equation (5) is obtained under the assumption that there is no hole degeneration in the base region of silicon near the $p^+$ contact. However, in the case when the surface charge concentration of acceptors $N$ exceeds $10^{12}$ cm$^{-3}$, holes may be degenerate in the near-surface region of the base region.

Then, the electroneutrality equation must be formulated taking degeneracy into account

$$N = \left(\frac{2kT\varepsilon_0\varepsilon_{Si}}{q^2}\right)^{1/2} [N_V F_{3/2}(y_{0w})]^{1/2}, \tag{1A}$$

where the Fermi-Dirac function is defined by

$$F_{3/2}(y_{0w}) = \frac{4}{3\sqrt{\pi}} \int_0^\infty \frac{\varepsilon^{3/2} d\varepsilon}{1 + \exp(y_{0w} + \varepsilon)N_V/p_0}. \tag{2A}$$

Here $N_V$ is the effective density of states in the valence band, and $p_0$ the equilibrium hole concentration in the bulk base region.

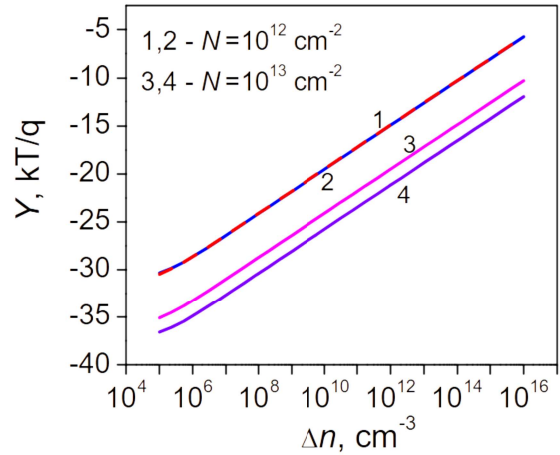

Fig. 1A. The non-equilibrium dimensionless potential $y_s$ on the surface of the silicon base region vs. excess concentration of electron-hole pairs $\Delta n$ at $N = 10^{12}$ (curves 1, 2) and $N = 10^{13}$ cm$^{-3}$ (curves 3, 4). Curves 1 and 3 are calculated without and curves 2 and 4 with taking degeneracy into account.

Plotted in Fig. 1A is the dependence of the non-equilibrium dimensionless potential $y_s$ on the surface of the silicon base region bordering with the $p^+$ contact on the excess concentration of electron-hole pairs $\Delta n$ for the cases when the value of $N$ is $10^{12}$ cm$^{-2}$ and $10^{13}$ cm$^{-3}$, using equations (5), (1A) and (2A). As can be seen from the figure, the $y_s(\Delta n)$ curves, calculated with and without taking into account degeneracy at $N = 10^{12}$ cm$^{-2}$ (curves 1 and 2), practically coincide, which indicates the absence of degeneracy. As shown by the calculation of the value of the surface concentration of holes at the interface with the p$^+$ contact, it is about $3 \cdot 10^{18}$ cm$^{-3}$, which is almost an order of magnitude smaller than the value of $N_V$, which at room temperature is $1.8 \cdot 10^{19}$ cm$^{-3}$. In Fig. 1A the dependences of the non-equilibrium dimensionless potential $y_s(\Delta n)$ are also plotted for the case when the value of $N$ is $10^{13}$ cm$^{-3}$. In this case, the curves plotted with (curve 3) and without (curve 4) consideration of degeneracy differ.

However, the values of the SCR recombination velocities obtained using equation (2) and their dependence on the excess concentration of charge carriers coincide both at $N = 10^{12}$ cm$^{-3}$ and at $N = 10^{13}$ cm$^{-3}$, both with and without taking into account degeneracy (see Fig. 2A). This is related to the fact that the depletion layer, in which the recombination times for holes and electrons either coincide or differ not very much, makes the greatest contribution to SCR recombination.

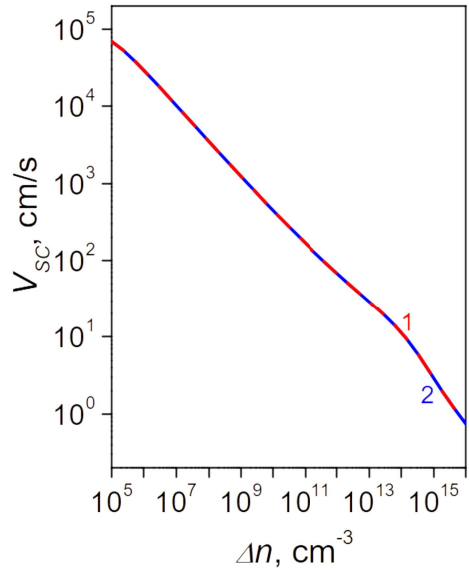

Fig. 2A. SCR recombination velocity vs. excess concentration, obtained using Eqs. (5) та (1A).

In the general case, to find the coefficients $b_{wd}$ and $b_{w0}$, it is necessary to solve the generation-recombination equation, taking into account that the generation term of the equation is proportional to $e^{-\alpha x}$, where $\alpha$ is the absorption coefficient. When using all the equations given in the main text of the work, taking into account the specified term, we obtain the following expression for $b_{wd}(\alpha)$:

$$b_{wd}(\alpha) = \frac{2 + 2\left(\dfrac{S_{SC}}{\alpha D} - 1\right)\cosh\left(\dfrac{d}{L}\right)e^{-\alpha d} - \left(\dfrac{1}{\alpha L}\dfrac{S_{SC}L}{D} - 1\right)e^{\frac{-d}{L}-\alpha d} - \left(\dfrac{1}{\alpha L}\dfrac{S_{SC}L}{D} + 1\right)e^{\frac{d}{L}-\alpha d}}{2\left(\dfrac{S_{SC}}{\alpha D} - 1\right)e^{-\alpha d} + 2\cosh\left(\dfrac{d}{L}\right) + 2\dfrac{S_{SC}L}{D}\sinh\left(\dfrac{d}{L}\right) - \left(\dfrac{1}{\alpha L}\dfrac{S_{SC}L}{D} - 1\right)e^{-\frac{d}{L}} - \left(\dfrac{1}{\alpha L}\dfrac{S_{SC}L}{D} + 1\right)e^{\frac{d}{L}}}.$$

(3A)

It should be noted that the quantum efficiency of the short-circuit current in textured silicon SCs is determined by an expression of the type

$$q(\lambda, d, b) = \frac{\alpha(\lambda)}{1 + \dfrac{b}{4n_r^2(\lambda)d}},$$

(4A)

and not by

$$q(\lambda, d) = 1 - \exp(-\alpha(\lambda)d).$$

(5A)

Therefore, when obtaining the dependences of $b_{wd}(\alpha)$ for textured silicon SCs, it is necessary to select an effective value of $\alpha$, which allows to achieve agreement with experiment based on the use of quantum efficiency (5A) instead of (4A) when finding the short-circuit current in AM1.5 conditions. As the estimates show, the so obtained typical value of the absorption coefficient $\alpha$ in textured silicon SCs with an efficiency of the order of 20% or higher is approximately $1.1 \cdot 10^4$ cm$^{-1}$. In Fig. 3A are shown the dependences of $b_{wd}(\tau_R)$ for the cases when the value of $\alpha$ is close to zero (at the same time, the generation term in the original equation is a constant), when the value of $\alpha = 1.1 \cdot 10^4$ cm$^{-1}$, and according to formula (15) for $b_{dw}$.

As the calculations show, in the case when $\alpha \approx 0$, the value of $b_{wd}(\tau_R)$ is greater than the value of $b_{dw}$, obtained according to Eq. (15), but in the case when $\alpha = 1.1 \cdot 10^4$ cm$^{-1}$, the values of $b_{wd}(\tau_R)$ and $b_{dw}$ match The latter case practically corresponds to the surface generation of light in the SC.

Let us further consider the important question about the concentration of deep impurities responsible for the SCR recombination in those SCs whose parameters are used in Figs. 6-8. By definition, in this case the valid ratio is $\tau_R = (C_p N_t)^{-1}$. Let us rewrite it in the following form

$$N_t = \frac{\tau_R^{-1}}{C_n b_r}.$$

(6A)

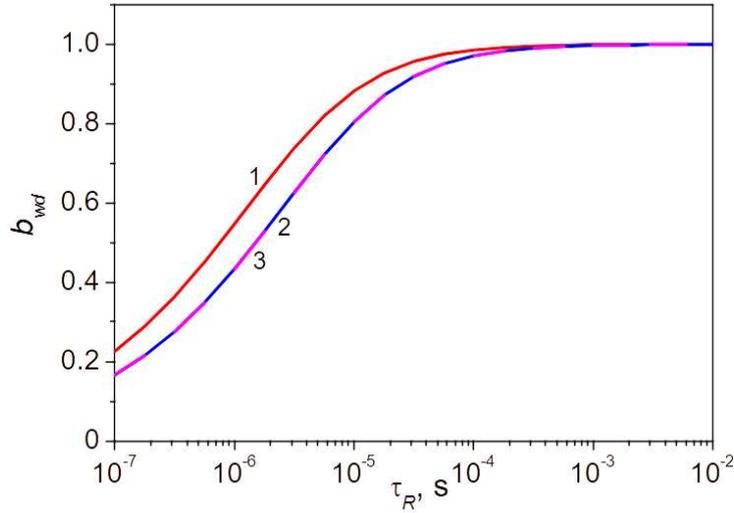

Fig.3A. The dependence of $\tau_R$ on $\Delta n$ obtained using Eqs. (3A) and (15).

Assume that the value of $C_n$ is constant and is equal to $10^{-9}$ cm$^3$/c. Then, substituting the values of $\tau_{Rm}$ and $b_r$, for which Figs. 6-8 were built, we obtain, respectively: $N_{tm1} = 1.2 \cdot 10^{15}$ cm$^{-3}$, $N_{tm2} = 8 \cdot 10^{13}$ cm$^{-3}$, $N_{tm3} = 7.14 \cdot 10^{14}$ cm$^{-3}$. If we recalculate the obtained values of deep-level concentrations for the studied SCs using their Gaussian distribution in the SCR, we get $N_1 = 4 \cdot 10^{10}$ cm$^{-2}$, $N_2 = 7.3 \cdot 10^9$ cm$^{-2}$, $N_3 = 8.1 \cdot 10^9$ cm$^{-2}$, respectively. Despite the fact that the maximum concentrations of deep levels for different SCs differ by more than an order of magnitude, the integral concentrations differ by no more than one and a half times. It is also surprising that the smallest concentration of levels occurs for the SC, in which the value of $\tau_R$ is the smallest.

Finally, we discuss the question about the criteria that should be used when calculating the SCR recombination velocity based on either the $\tau_R$ model or the model with a Gaussian distribution of centers in the SCR. The approximation $\tau_R \approx$ const can be used when the integrand in Eq. (1) fits completely into the Gaussian. This case is realized in the region of small and medium excitation levels, as a rule, when $\Delta n < 10^{14}$ cm$^{-3}$. If the integrand function in (1) goes beyond the Gaussian, then the general formula with the Gaussian (23) should be used. This case is realized at sufficiently high excitation levels, when $\Delta n > 10^{14}$ cm$^{-3}$.

**Appendix 2**

The recombination component associated with the non-radiative recombination of excitons on bulk recombination centers by the Auger mechanism is described by expression (38). Let's go back to the history of the issue of this recombination. In papers [23-26] it was shown that in semiconductors, in particular in silicon, at sufficiently large values of Δ$n$, when Δ$n$ > $n_0$, there are two subsystems, electron-hole and exciton. Recombination, both radiative and non-radiative, occurs both through the electron-hole and exciton channels. There is interaction between these subsystems.

In [23], it was established that depending on the value of the excitation level, there are cases when the presence of excitons can be neglected (low excitation level), when excitons affect the effective lifetime of electron-hole pairs (intermediate excitation), and when excitons determine all characteristics (high level of excitation).

Does the presence of an exciton channel affect the characteristics of silicon solar cells? At the excess charge carrier concentration of the order of $10^{15}$ cm$^{-3}$, which correspond to the point of maximum power selection, the concentration of excitons, which is proportional to the product *pn*, is significantly smaller than the value of Δ$n$. It would seem that excitons should not make a significant contribution to the total recombination. But, as was shown in the works of Hangleiter [27,28], the spatial localization of an electron and a hole in an exciton significantly increases the probability of Auger processes, both in the form of direct band-to-band recombination and in the form that involves impurity centers. Therefore, in this case, an intermediate level of excitation is realized. Both the first and the second cases were analyzed in works [27, 28]. At the same time, it was established that the same deep level can provide both recombination according to the Shockley-Reed-Hall mechanism and Auger exciton recombination involving deep centers. In [29], the contribution of excitons to the band-to-band recombination in silicon was taken into account, whereas the work [21] incorporates the contribution of excitons to Auger recombination with the participation of deep impurities. It turned out that the effective bulk lifetime of charge carriers in silicon, taking into account this contribution, depends on the level of doping according to

$$\tau_{eff}(n_0) = \frac{\tau_{SRH}}{1+\dfrac{n_0}{8.2\cdot 10^{15}}} \ . \tag{1B}$$

Well before the paper [21], a work [30] was published, in which a similar empirical expression for the dependence of the effective bulk lifetime on the level of doping was proposed, in which the value 7.1·10$^{15}$ cm$^{-3}$ appeared instead of 8.2·10$^{15}$. In work [30], the proposed expression was compared for a large number of samples with both electron and hole conductivity and a good agreement between empirical and experimental data was obtained. The difference between the characteristic concentration values has a simple explanation: In the work [21], the obtained dependence was isolated from the interband recombination, whereas in [30] this was not done.

In silicon samples with long Shockley-Reed-Hall lifetimes (greater than a millisecond), the non-radiative exciton recombination mechanism with the lifetime (1A) works together with the Shockley-Reed-Hall channel. Thus, at $\tau_{SRH}$ of the order of 10 ms, the time of non-radiative exciton recombination significantly reduces their total value in the range of doping levels from 10$^{15}$ to 10$^{16}$ cm$^{-3}$. In samples with the doping concentration of 10$^{15}$ cm$^{-3}$, this decrease is 12%, and in samples with the doping level of 4.9·10$^{15}$ cm$^{-3}$ it is 60%.

The lifetime of non-radiative exciton recombination is structurally closest to the time of radiative recombination

$$\tau_r^{-1} = B_r \left( n_0 + \Delta n \right) , \qquad (2B)$$

where $B_r$ is radiative recombination coefficient. Calculations and comparison of the lifetime of non-radiative exciton recombination and the lifetime of radiative recombination in silicon show that at $\tau_{SRH} \leq 50$ ms the former time is shorter than the latter. Even in the record-breaking efficiency of silicon SCs, the value of $\tau_{SRH}$ is of the order of 10 ms. Therefore, it is logical, if one takes into account radiative recombination, to also take into account non-radiative exciton recombination.

To confirm this statement, in Fig. 1B is shown the dependences of the effective bulk lifetime for n-Si samples with long Shockley-Reed-Hall lifetimes on the doping level, taken from Richter's work [20]. Assuming that expression (1B) is valid and that for this reason the experimental points in the range from 10$^{15}$ to 5.3·10$^{15}$ cm$^{-3}$ are lower, we obtained corrected values that take into account only the influence of radiative recombination and band-to-band recombination, by multiplying the given values by the factor (1 + $n_0$/8.2·10$^{15}$ cm$^{-3}$) (blue curve). The use of theoretical dependences that take into account Shockley-Reed-Hall recombination, nonradiative exciton recombination, radiative recombination, and band-to-band Auger recombination made it possible to reconcile the corrected experiment with theory. At the same time, the calculation, which does not take into account the non-radiative exciton recombination, does not agree with the experiment (see the red curve).

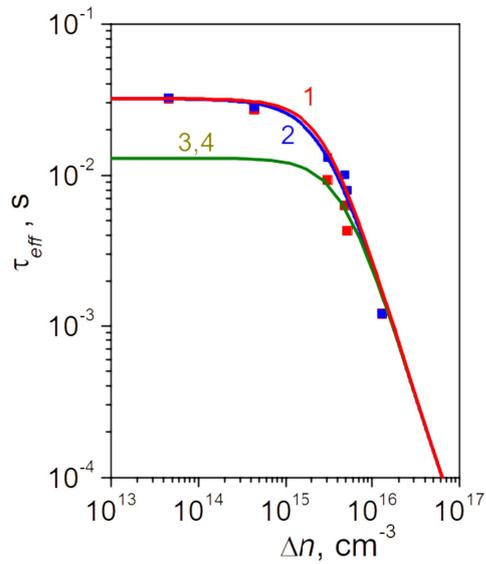

Fig. 1B. Experimental effective lifetime $\tau_{eff}$ in silicon vs the doping level, taken from work [20] (symbols) and theoretical dependences obtained with (curves 1 and 2) and without (curves 3 and 4) the non-radiative exciton recombination (lines).

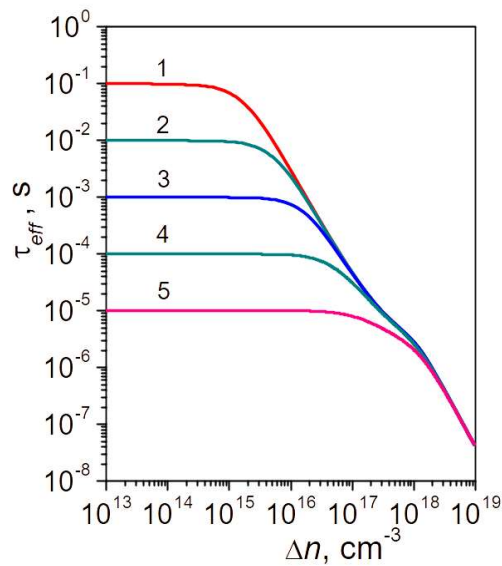

Fig. 2B. Theoretical effective lifetime $\tau_{eff}$ in silicon on the doping level at different Shockley-Reed-Hall lifetimes $\tau_{SRH}$ = $10^{-1}$(1), $10^{-2}$(2), $10^{-3}$(3), $10^{-4}$( 4) and $10^{-5}$ s (5), obtained taking into account the lifetime of nonradiative exciton recombination (lines).

lifetime is determined by the Shockley-Reed-Hall mechanism and the non-radiative exciton recombination time. So, for example, at a value of $\tau_{SRH}$ equal to $10^{-5}$ s, this range continues up to the values of $n_0$ that exceed $10^{17}$ cm$^{-3}$.

It should be noted that in [25] excitonic nonradiative recombination was actually taken into account by using the lifetime given in [29].

Thus, the excitonic nonradiative recombination must always be taken into account as a separate recombination channel.